\documentclass[10pt,english,eqsecnum,english,prd,aps,nofootinbib,superscriptaddress,amsfonts,longbibliography,showpacs]{revtex4-1}
\usepackage[T1]{fontenc}
\usepackage[latin9]{inputenc}
\usepackage{color}
\usepackage{babel}
\usepackage{units}
\usepackage{mathrsfs}
\usepackage{amsmath}
\usepackage{amssymb}
\usepackage{graphicx}
\usepackage{esint}
\usepackage[unicode=true,pdfusetitle,
 bookmarks=true,bookmarksnumbered=false,bookmarksopen=false,
 breaklinks=false,pdfborder={0 0 1},backref=false,colorlinks=false]
 {hyperref}
\usepackage{breakurl}

\makeatletter
\allowdisplaybreaks[1]

\makeatother

\begin{document}

\title{Polymer quantization and the saddle point approximation of partition functions}

\author{Hugo A. Morales-T\'{e}cotl}

\email{hugo@xanum.uam.mx}

\author{Daniel H. Orozco-Borunda}

\author{Saeed Rastgoo}

\email{saeed@xanum.uam.mx}

\selectlanguage{english}%

\affiliation{Departamento de F\'{i}sica, Universidad Aut\'{o}noma Metropolitana
- Iztapalapa\\
San Rafael Atlixco 186, Mexico D.F. 09340, Mexico}

\date{\today}

\begin{abstract}

The saddle point approximation of the path integral partition functions is an important
way of deriving the thermodynamical properties of black holes. However,
there are certain black hole models and some mathematically analog mechanical
models for which this method cannot be applied directly. This is due to the
fact that their action evaluated on a classical solution is not finite
and its first variation does not vanish for all consistent boundary conditions. These problems can be dealt with
by adding a counterterm to the classical action, which is a solution
of the corresponding Hamilton-Jacobi equation. 

In this work we study the effects of polymer quantization on a mechanical model presenting the aforementioned difficulties and contrast it with the above counterterm
method. This type of quantization for mechanical models is motivated by the loop quantization of gravity which is known to play a role in the thermodynamics
of black hole systems.

The model we consider is a nonrelativistic
particle in an inverse square potential, and analyze two polarizations
of the polymer quantization in which either the position or the momentum
is discrete. In the former case, Thiemann's regularization is
applied to represent the inverse power potential but we still need to incorporate the Hamilton-Jacobi counterterm
which is now modified by polymer corrections.
In the latter, momentum discrete case however, such regularization could not be implemented. Yet, remarkably, 
owing to the fact that the position is bounded, we do not
need a Hamilton-Jacobi counterterm in order to have a well-defined
saddle point approximation. Further developments and extensions are commented upon in the discussion.
\end{abstract}
\maketitle
%\tableofcontents{}

\section{Introduction}

Two major open problems in theoretical physics regard the nature of spacetime: on one hand there is the issue of singularities, beyond which classical general relativity cannot be continued, and, on the other, one finds the divergent high energy behavior of field theories. A quantum theory of gravity is expected to have a bearing on both of these problems. For instance, loop quantum gravity \cite{Rovelli:2004tv,Thiemann:2007zz,GambiniPullin2011,Bojowald2011,RovelliVidotto2014} has been shown indeed to replace the big bang of general relativistic homogeneous cosmological models by a bounce \cite{Bojowald:2001xe,Ashtekar:2006wn} and quantum field theory in such a scenario \cite{Thiemann:1997rt} is rather different from usual fixed background field theory. Moreover, it has also provided a specific account for black hole entropy \cite{Ashtekar:1997yu}.

Now to investigate the behavior of some simple systems under this approach it is possible to use polymer quantum mechanics, a finite number of degrees of freedom scheme including some of the loop quantum gravity techniques \cite{Ashtekar:2002sn}. This simplified approach has been applied to some systems to contrast their behavior with their usual Schr\"odinger quantization and its relation with the latter either as a continuum limit \cite{Corichi:2006qf,Corichi:2007tf} or as a low energy approximation \cite{Flores-Gonzalez:2013zuk}. Furthermore, for the case of fields on a fixed background this technique has been applied to each of its infinite modes as a candidate to explore their high energy behavior \cite{Hossain:2010eb,Garcia-Chung:2014sza}. Even higher order derivative models have been given consideration along these lines recently \cite{Cumsille:2015xaa}.
With the exception of \cite{HusainWinkler2003} that advanced a path integral version of polymer quantum mechanics, most work on these lines adopted a Hamiltonian scheme. This changed recently:  a path integral approach was considered \cite{Ashtekar:2009dn,Ashtekar:2010gz,Ashtekar:2010ve} in order to provide a more detailed link between loop quantum cosmology and the covariant spin foam models \cite{RovelliVidotto2014}, and a polymer path integral in quantum field theory and its relation to Lorentz invariance has also been studied in \citep{kajuri2014path}.  Also the Feynman formula for other mechanical examples has been worked out \cite{Parra:2014mxa},  and explicit polymer propagators have been obtained \cite{MoralesRuelas2015}.
An interesting aspect of this so-called Feynman approach is that semiclassical approximations can be at hand to investigate important gravitating systems through the saddle point approximation of its path
integral description. In fact, the Euclidean path integral is specially
useful in studying the thermodynamics of systems such as black holes
since there it is interpreted as the partition function of the system in
a canonical ensemble. 

Let us consider the semiclassical approximation for a system with the Euclidean path integral
\begin{equation}
\mathcal{Z}=\int\prod_{j}\mathscr{D}\phi_{j}\exp\left(-\frac{1}{\hbar}S_{E}[\phi_{j}]\right),\label{eq:path-int-gen}
\end{equation}
in which $\phi_{j}$ are the fields of the theory and $S_{E}$ is
its Euclidean action. Given that one can expand the action around the classical
solutions $\phi_{j}\big|_{cl}$ as
\begin{equation}
S_{E}\left[\left.\phi_{j}\right|_{cl}+\delta\phi_{j}\right]=S_{E}\left[\left.\phi_{j}\right|_{cl}\right]+\delta S_{E}\left[\left.\phi_{j}\right|_{cl}+\delta\phi_{j}\right]+\frac{1}{2}\delta^{2}S_{E}\left[\left.\phi_{j}\right|_{cl}+\delta\phi_{j}\right]+\cdots,\label{eq:expnd-SE}
\end{equation}
one can then substitute this into (\ref{eq:path-int-gen}); by keeping  only up to the quadratic term, one gets 
\begin{equation}
\mathcal{Z}\approx\exp\left(-\frac{1}{\hbar}S_{E}\left[\left.\phi_{j}\right|_{cl}\right]\right)\int\prod_{j}\mathscr{D}\phi_{j}\exp\left(-\frac{1}{2\hbar}\delta^{2}S_{E}\left[\left.\phi_{j}\right|_{cl}+\delta\phi_{j}\right]\right),\label{eq:saddle-gen}
\end{equation}
which is the saddle point approximation to the model that gives us
access to the semiclassical information about the system.

More precisely, a saddle point approximation (\ref{eq:saddle-gen}) of (\ref{eq:path-int-gen})
is possible only if the following conditions are met:
\begin{enumerate}
\item The variational
principle is well-defined: $S_{E}$ must be functionally differentiable for all the variations
of the fields and compatible with the boundary and falloff conditions of the fields, so that any boundary term must vanish by virtue of these conditions;  thus we can write $\delta S_{E}=\int d^{n}x\frac{\delta S_{E}}{\delta\phi_{j}}\delta\phi_{j}$. This is needed so that one is able to

\begin{enumerate}
\item expand $S_{E}$ as in (\ref{eq:expnd-SE}) around an arbitrary configuration
$\phi_{0}$ using the functional derivatives,
\item find the classical solutions by setting $\delta S_{E}=0$,
\item perform the expansion (\ref{eq:expnd-SE}) specifically around the classical solutions $\left.\phi_{j}\right|_{cl}$.
\end{enumerate}
\item Given condition 1, then $S_{E}$ on classical solutions must remain finite,
i.e., $S_{E}\big|_{\delta S_{E}=0}<\infty$.
\item Given conditions 1 and 2, the Gaussian integral  $\int\prod_{j}\mathscr{D}\phi_{j}\exp\left(-\frac{1}{2\hbar}\delta^{2}S_{E}\left[\left.\phi_{j}\right|_{cl}+\delta\phi_{j}\right]\right)$
also must remain finite.
\item Since there is a minus sign in the exponent of (\ref{eq:path-int-gen}), we should
also have $\delta^{2}S_{E}\left[\left.\phi_{j}\right|_{cl}+\delta\phi_{j}\right]>0$
so that the classical solutions give the dominant contribution to
the saddle point approximation.
\end{enumerate}

The physical reason behind the above conditions is that for a semiclassical
approximation, the most important contribution comes from the classical
solutions, and thus everything is expanded around
such a trajectory. The rest of the terms become less and less important, and
thus we only need to keep the perturbative terms up to the quadratic term,
which gives us the nonclassical contributions. 
%The mathematical argument behind the conditions is that the most important contribution to the exponential $\exp\left(-\frac{1}{\hbar}S_{E}[\phi_{j}]\right)$, and hence to the integral, comes from the minimum of $S_{E}[\phi_{j}]$ (due to the minus sign in the exponent of $\exp$) which happens when $\delta S_{E}=0$ and $\delta^{2}S_{E}>0$.

There are, however, some important systems for which these conditions
are not met and thus access to the semiclassical approximation via
the saddle point method is not possible \citep{Regge1974,Gibbons1977,H.Liebl1997,Papadimitriou2005,McNees2005,Mann2006}.
Among these systems are a class of two-dimensional (2D) dilatonic gravitational systems,
including models like CGHS (Callan-Giddings-Harvey-Strominger model) \citep{C.G.Callan1992} and 3+1 spherically
symmetric,  which have black hole solutions and many other related
interesting properties. This class can be described by the generic
action \citep{Grumiller2007} 
\begin{equation}
S=-\frac{1}{16\pi G_{2}}\int_{\mathcal{M}}d^{2}x\sqrt{-g}\left[\Phi R-U(\Phi)\nabla_{a}\Phi\nabla^{a}\Phi-2V(\Phi)\right]-\frac{1}{8\pi G_{2}}\int_{\partial\mathcal{M}}dx\sqrt{q}\Phi K.\label{eq:dilaton-act}
\end{equation}
Here $\Phi$ is the dilaton field, and $U(\Phi)$ and $V(\Phi)$ are
model-dependent functions of the dilaton field. The latter is called
the dilaton potential. The boundary term is the equivalent of the
Gibbons-Hawking-York term \citep{York1972} in this theory in which
$K$ is the trace of the extrinsic curvature of the boundary manifold.

To see the problem with saddle point approximation, we consider the
on-shell variation of the action \citep{Grumiller2007}
\begin{equation}
\delta S\big|_{cl}=\int_{\partial\mathcal{M}}dx\sqrt{q}\left(\Xi^{ab}\delta q_{ab}+\Upsilon_{\Phi}\delta\Phi\right).
\end{equation}
It can be shown that this variation does not vanish in some of the models
of this generic class. This is basically due to the fact that in these
models, the coefficients of the variations of the spatial metric $q_{ab}$
and the dilaton $\Phi$ in the above expression, diverge more rapidly
than the variations themselves fall off. This might look odd since
the presence of the Gibbons-Hawking-York  (GHY) term is supposed to
guarantee the well-posedness of the variational principle in these
kinds of theories. Notice, however, that the presence of the GHY terms is
 to let us avoid prescribing Neumann boundary conditions for the metric, i.e. it cancels
all the variations $\delta(\partial_{a}g_{bc})$ that come from the
bulk term. It does not guarantee that the Dirichlet boundary
conditions on the dilaton field lead to a well-posed variational principle. Also it may not be helpful in dealing with issues emanating from falloff conditions of the dilaton field. It is these types of boundary and falloff conditions that contribute to the problem here.

In some of the submodels
of this class, even if the variational principle is well defined, the on-shell value of the action diverges, especially due to the falloff conditions on the dilaton field
and its value on the boundary. So the saddle point approximation collapses
for these types of models. 

Since as mentioned above, one important use of saddle point approximation
is to study the thermodynamics of black holes, not being able to make
such an approximation for this class of 2D models is a significant
shortcoming. Luckily there is a common and rather generic method of
fixing this problem that consists of the addition of a boundary counterterm to
$S_{E}$, which not only makes the action functionally differentiable for all boundary and falloff conditions, but also renders the on-shell value of the action finite. It turns out that this boundary counterterm is a solution to the Hamilton-Jacobi equation of the
system \citep{Grumiller2007}. The action that is the sum of (\ref{eq:dilaton-act})
and the Hamilton-Jacobi boundary term is called the improved action.
An interesting observation is that the improved action actually gives
the correct thermodynamics for the system \citep{Grumiller2007}. 

In this paper, we investigate the behavior of the polymer path integral quantization of a mechanical model that in the usual path integral quantization suffers from the aforementioned ill-defined  semiclassical approximation. Whether this fixes the problem, and whether we need a Hamilton-Jacobi counterterm, it is relevant to know how the polymer approach changes the partition function for such systems and, perhaps, the thermodynamics of some of the 2D black holes. The present work is a first
step in this direction, and we study a simpler analog model of the aforementioned 2D class, namely, a particle in an inverse square potential, that has the same technical problems within the usual nonpolymer path integral quantization. Let us notice that the Hamiltonian polymer quantization of a particle subject to a Coulomb 
\citep{Husain2007} and an inverse squared \cite{Kunstatter:2008qx} potential, respectively, have been considered previously. Hence, our work complements these studies.

The paper is organized as follows. In Sec. \ref{sec:analog-model},
we introduce the analog model and describe the saddle point issue
for this case. Then we recall its solution through
the application of the Hamilton-Jacobi method. In Sec.
\ref{sec:eff-poly-with-HJ}, we polymerize this analog model in a
polarization where the position $q$ is discrete. We will see that
in this polarization, one still needs to add a counterterm to the
action to get a new well-defined action suitable for saddle point
approximation. However, we show that this counterterm and the
bulk action are both modified by polymer quantization. In Sec.
\ref{sec:eff-poly-without-HJ} we propose an effective potential for
the semiclassical version of the analog model in a polarization with
discrete momentum $p$. We then describe how this effective
potential does not require the addition of any counterterm
to the action. Finally we summarize and make our concluding remarks
in Sec. \ref{sec:conclude}. Details of the calculations
are given in the appendixes at the end of the paper.

\section{Analog mechanical model and its improved action\label{sec:analog-model}}

There are several types of simpler analog models that exhibit the aforementioned ill-defined semiclassical approximation. One such class of models corresponds to  single particle systems
in half-binding potentials $W$ \citep{Grumiller2007a}. We choose
a simple system in this class with an inverse square potential with
the Newtonian action (the subscripts $N$ and $E$ stand for Newtonian
and Euclidean, respectively)
\begin{equation}
S_{N}=\int_{t_{i}=0}^{t_{f}=\infty}dt\,L_{N}=\int_{t_{i}=0}^{t_{f}=\infty}dt\left(\frac{m}{2}\left(\frac{dq}{dt}\right)^{2}-W(q)\right),\,\,\,\,\,\,\,\,\,\,\,\,\,\,\,\,\,\,\,\,\,\,\,W(q)=\frac{k}{q^{2}}\label{eq:Lorentz-action}
\end{equation}
and corresponding Euclidean action (by a Wick rotation $t\rightarrow\tau=it$), $S_N=iS_E$,
\begin{equation}
S_{E}=\int_{\tau_{i}=i0}^{\tau_{f}=i\infty}d\tau\,L_{E}=\int_{\tau_{i}=i0}^{\tau_{f}=i\infty}d\tau\left(\frac{m}{2}\left(\frac{dq}{d\tau}\right)^{2}+W(q)\right).\label{eq:Euclid-action}
\end{equation}
The Newtonian equation of motion is
\begin{equation}
m\frac{d^{2}}{dt^{2}}q=-\frac{\partial W}{\partial q}\left(=\frac{2k}{q^{3}}\right),\label{eq:Lorentz-EOM}
\end{equation}
which under a Wick rotation becomes
\begin{equation}
m\frac{d^{2}}{d\tau^{2}}q=\frac{\partial W}{\partial q}\left(=-\frac{2k}{q^{3}}\right).\label{eq:Euclid-EOM}
\end{equation}
Let us see the problems of the saddle point approximation
for this analog model. First, we consider the variation of the Euclidean action
\begin{equation}
\delta S_{E}=  \left[\frac{\partial L_{E}}{\partial\frac{d}{d\tau}q}\delta q\right]_{0}^{\infty}+\int_{0}^{\infty}d\tau\left(-\frac{d}{d\tau}\frac{\partial L_{E}}{\partial\frac{d}{d\tau}q}+\frac{\partial L_{E}}{\partial q}\right)\delta q.\label{eq:delta-S-bndry}
\end{equation}
If the boundary term does not vanish for all the variations
of $\delta q$ compatible with the boundary and falloff conditions, then the action is not functionally differentiable. It turns out that this is the case. At $\tau=it_{i}=i0$, since
the value of $q$ is finite, we have $\delta q=0$. However, at $\tau=it_{f}=i\infty$,
since $q\rightarrow\infty$, the condition $\delta q \big|_{\infty} \neq 0$ is also allowed; i.e., any two trajectories do not necessarily coincide at infinity, and yet they both tend to infinite values. Thus the action is not functionally differentiable with this boundary condition.

The common way to overcome this issue is to add a boundary term $-B$ to the action that cancels out the boundary term present in (\ref{eq:delta-S-bndry}). Clearly the variation of the boundary term should obey $\delta B=\frac{\partial L}{\partial\frac{d}{d\tau}q}\delta q=p\delta q$.
It just happens that Hamilton's principal function $G$ has exactly
this property. This is because we have
\begin{equation}
\delta G(q,t)=\frac{\partial G}{\partial q}\delta q=p\delta q=\frac{\partial L}{\partial\frac{d}{d\tau}q}\delta q.
\end{equation}
Hence we conclude that by adding 
\begin{equation}
-B=-G
\end{equation}
to the action, it becomes functionally differentiable. Such an action is called the ``improved'' action, 
\begin{equation}
\Gamma[q]=S_{E}-\left.G(q,t)\right|_{0}^{t_{f}}=\int_{0}^{t_{f}}dt\left(\dfrac{m\dot{q}^{2}}{2}+\dfrac{k}{q^{2}}\right)-\left.G(q,t)\right|_{0}^{t_{f}}.\label{eq:imprv-act-gen}
\end{equation}
Clearly, the variation of (\ref{eq:imprv-act-gen}) yields
\begin{equation}
\delta S_{E}=\int_{0}^{\infty}d\tau\left(-\frac{d}{d\tau}\frac{\partial L_{E}}{\partial\frac{d}{d\tau}q}+\frac{\partial L_{E}}{\partial q}\right)\delta q+\left[\left(\frac{\partial L_{E}}{\partial\frac{d}{d\tau}q}-\frac{\partial G}{\partial q}\right)\delta q\right]_{0}^{\infty}=\int_{0}^{\infty}d\tau\left(-\frac{d}{d\tau}\frac{\partial L_{E}}{\partial\frac{d}{d\tau}q}+\frac{\partial L_{E}}{\partial q}\right)\delta q.\label{eq:imprv-eval}
\end{equation}
Next, we consider the Euclidean action itself. We would like to show that even if the action is functionally differentiable, the value of the Euclidean action on classical solutions is not finite, and that the addition of $G$ to the action, makes it finite. Assuming for the moment that $S_E$ is functionally differentiable, using the Leibniz rule on the kinetic term and
then computing the action on the classical solutions, we can write
\begin{equation}
S_{E}\big|_{q_{cl}}=\left[\frac{m}{2}q_{cl}\frac{dq}{d\tau}\bigg|_{cl}\right]_{\tau=i0}^{\tau=i\infty}+\int_{i0}^{i\infty}d\tau\left(\left.-\frac{m}{2}q_{cl}\frac{d^{2}q}{d\tau^{2}}\right|_{cl}+\frac{k}{q_{cl}^{2}}\right).\label{eq:SE-Leib}
\end{equation}
Using (\ref{eq:Euclid-EOM}) and the form of the classical solution
$q_{cl}(\tau)$, the integral turns out to be finite and of order 
$\pi$. On the other hand, it can clearly be seen that the boundary
term above (or its Newtonian counterpart) diverges, since at $\tau\rightarrow\infty$
(or $t\rightarrow\infty$) the system behaves as a free particle, 
$\frac{dq}{d\tau}|_{cl}$ (or $\frac{dq}{dt}|_{cl})$ is constant,
and $q_{cl}\rightarrow\infty$. Thus the action evaluated on classical
solutions is not finite, 
\begin{equation}
S_{E}\big|_{q_{cl}}\rightarrow\infty.\label{eq:S-inf}
\end{equation}
These results show that the direct saddle point approximation cannot be performed for this system. 

Now let us see what is the form of $G$ and how it makes the action, evaluated on classical solutions, finite. The variable $t$ is used from now on, both as Euclidean or Newtonian time, depending on the context. Since  $G$ in (\ref{eq:imprv-act-gen})
is a solution of the Hamilton-Jacobi equation,
we have\begin{equation}
H\left(q,\dfrac{\partial G}{\partial q}\right)+\dfrac{\partial G}{\partial t}=0.\label{eq:HJ-eq-gen}
\end{equation}
To find the explicit form of $G$ we need to solve (\ref{eq:HJ-eq-gen}).
Considering the Euclidean Hamiltonian
\begin{equation}
H=\frac{p^{2}}{2m}-\frac{k}{q^{2}},
\end{equation}
and noting that $p=\frac{\partial G}{\partial q}$, the Eq. (\ref{eq:HJ-eq-gen})
becomes
\begin{equation}
\frac{1}{2m}\left(\frac{\partial G}{\partial q}\right)^{2}-\frac{k}{q^{2}}+\frac{\partial G}{\partial t}=0.\label{eq:HJ-inv-pot}
\end{equation}
Since the Hamiltonian does not depend explicitly on $t$, Eq. (\ref{eq:HJ-eq-gen})
implies that the derivative $\nicefrac{\partial G}{\partial t}$ must
be a constant in time and is actually the negative of the energy
of the system $H=E$. So one can make an additive separation of the variables
in $G$ as
\begin{equation}
G(q,E,t)=\Omega(q,E)-Et,\label{eq:G-w-E}
\end{equation}
which turns (\ref{eq:HJ-inv-pot}) into 
\begin{equation}
\frac{1}{2m}\left(\frac{\partial\Omega}{\partial q}\right)^{2}-\frac{k}{q^{2}}=E\label{eq:HJ-no-t-deriv}
\end{equation}
or
\begin{equation}
\left(\frac{\partial\Omega}{\partial q}\right)=\frac{\sqrt{2m}\sqrt{Eq^{2}+k}}{q}.\label{eq:w-in-HJ}
\end{equation}
This can be integrated to yield $\Omega$, which together with (\ref{eq:G-w-E})
leads to
\begin{equation}
G(q,t)=c_{0}-Et+\sqrt{2}\sqrt{m}\sqrt{Eq^{2}+k}-\sqrt{2}\sqrt{mk}\tan^{-1}\left(\sqrt{\frac{k}{Eq^{2}+k}}\right).\label{eq:G-explic}
\end{equation}
Since at $q\rightarrow\infty$ the potential is zero, the energy takes  the  form $E=\frac{1}{2}m\frac{q^{2}}{t^{2}}$ of a free particle
(which is equal to $p^{2}/2m$ asymptotically), thus  Hamilton's principal function behaves asymptotically like
\begin{equation}
G\approx\frac{m}{2}\frac{q^{2}}{t}.
\end{equation}
Noticing that asymptotically $\dot{q}=\nicefrac{q}{t}$ and hence $G\rightarrow\frac{m}{2}q\dot{q}$, the boundary term in the improved
action using (\ref{eq:SE-Leib}) and (\ref{eq:imprv-act-gen}) vanishes,
\begin{equation}
\left[\frac{m}{2}q_{cl}\dot{q}\bigg|_{cl}-G\right]_{0}^{\infty}=0,
\end{equation}
thus making $\Gamma[q_{cl}]$  finite. It is also worth noting that (\ref{eq:G-explic}) leads to 
\begin{equation}
\frac{\partial G}{\partial q}=\frac{\sqrt{2m}\sqrt{Eq^{2}+k}}{q^{2}},
\end{equation}
and since asymptotically $E\mid_{t_{f}\rightarrow\infty}=\frac{p^{2}}{2m}$, one gets 
\begin{equation}
\left.\frac{\partial G}{\partial q}\right|_{q\rightarrow\infty}\approx p,
\end{equation}
which is an explicit way of seeing $\delta\Gamma[q_{cl}]=0$.
 
Next we analyze how the polymer quantization affects the above argument. For this we consider two polarizations
of the polymer representation for the analogue model in the following sections.

\section{The polymer model with discrete position\label{sec:eff-poly-with-HJ}}

Polymer quantum mechanics is based on the idea of using the polymer
representation of the Weyl algebra, a singular representation that does not
obey the Stone-von Neumann theorem and hence is not equivalent to Schr\"{o}dinger representation \citep{Corichi:2007tf}. As is
well known, the Weyl algebra is based on the Weyl relations 
\begin{align}
\hat{U}_{\mu_{1}}\hat{U}_{\mu_{1}}= & \hat{U}_{\mu_{1}+\mu_{2}},\\
\hat{V}_{\lambda_{1}}\hat{V}_{\lambda_{2}}= & \hat{V}_{\lambda_{1}+\lambda_{2}},\\
\hat{U}_{\mu}\hat{V}_{\lambda}= & e^{-\frac{i}{\hbar}\lambda\mu}\hat{V}_{\lambda}\hat{U}_{\mu},
\end{align}
with the reality conditions $\hat{U}_{\mu}^{\dagger}=\hat{U}_{-\mu}$
and $\hat{V}_{\lambda}^{\dagger}=\hat{V}_{-\lambda}$. In the polymer
representation of this algebra, the corresponding Hilbert space, $\mathcal{H}_{\textrm{poly}}$,
possesses an uncountable orthonormal basis such that 
\begin{equation}
\langle\mu|\nu\rangle=\delta_{\mu,\nu},\,\,\,\,\,\,\,\,\,\,\mu,\nu\in\mathbb{R},
\end{equation}
where $\delta_{\mu,\nu}$ is a Kronecker delta. 

One can consider two polarizations of this polymer representation
in which either the representation of $\hat{U}_{\mu}$ or $\hat{V}_{\lambda}$
on $\mathcal{H}_{\textrm{poly}}$ is not weakly continuous in its
corresponding parameter, $\mu$ or $\lambda$. More precisely, by saying,
e.g., the representation of $\hat{V}_{\lambda}$ is not weakly continuous
with respect to $\lambda$, we mean $\lim_{\lambda\rightarrow0}\langle\mu|\hat{V}_{\lambda}|\mu\rangle\neq\langle\mu|\hat{V}_{\lambda=0}|\mu\rangle$.
A similar criterion applies for a polarization in which the representation
of $\hat{U}_{\mu}$ is not weakly continuous with respect to $\mu$.
Now, in a polarization where the representation of $\hat{V}_{\lambda}$
is not weakly continuous, the basic operators $\hat{U}_{\mu},\hat{V}_{\lambda}$
act on the basis vectors $|q\rangle$ as 
\begin{align}
\hat{U}_{\mu}|q\rangle= & e^{\frac{i}{\hbar}\mu q}|q\rangle,\label{eq:U-mu-prime}\\
\hat{V}_{\lambda}|q\rangle= & |q-\lambda\rangle.\label{eq:V-mu-prime}
\end{align}
Since $\hat{U}_{\mu}$ is weakly continuous in $\mu$ in this polarization,
one can write $\hat{U}_{\mu}=e^{\frac{i}{\hbar}\mu\hat{q}}$. However,
this is not the case for $\hat{V}_{\lambda}$. Namely, since the representation
of $V_{\lambda}$ is not weakly continuous with respect to $\lambda$,
the momentum $p$ cannot be represented on the Hilbert space $\mathcal{H}_{\textrm{poly}}$
in a well-defined manner as the generator of $\hat{V}_{\lambda}$.
Thus, although classically one may write $V_{\lambda}=e^{\frac{i}{\hbar}\lambda p}$,
this is not the case quantum mechanically and $\hat{V}_{\lambda}$
should be seen as an operator on its own and not as the exponentiation
of the generator $\hat{p}$. 
%This is what we mean by saying that the polymer representation is singular. 

Furthermore, since we cannot take the limit $\lambda\rightarrow0$
due to the singularity of the representation, and once we consider $\lambda$
as a fixed free parameter of the theory, one can see from \eqref{eq:V-mu-prime}
that by starting from a certain $q=q_{0}$, states get restricted
to a (one-dimensional) lattice in $q$ space where the wave functions
$\Psi(q)=\langle q|\Psi\rangle$ have nonvanishing values only on
the lattice points $\{q_{n}|q_{n}=q_{0}+n\lambda,\,n\in\mathbb{Z}\}$
for that $q_{0}$; here we choose $q_0=0$. Thus we say $q$ is discrete in this polarization
and write the basis $|q_{n}\rangle$ as a countable one, labeled with
$n$ and the value of the momentum is restricted to $-\frac{\pi\hbar}{\lambda}\leq p< \frac{\pi\hbar}{\lambda}$. We call this polarization with $q$ discrete,
$q$ polarization. The corresponding Hilbert space ${\mathcal H}_{q_0}$ is only a superselected sector of $\mathcal{H}_{\textrm{poly}}$, such that $\mathcal{H}_{\textrm{poly}} = \bigoplus_{0\leq q_0< \lambda} {\mathcal H}_{q_0}$.

In another polarization, one in which $\hat{U}_{\mu}$ is not weakly
continuous, things are the other way around: while we can write $\hat{V}_{\lambda}=e^{\frac{i}{\hbar}\lambda\hat{p}}$
(and also classically $U_{\mu}=e^{\frac{i}{\hbar}\mu q}$), we may
not write $\hat{U}_{\mu}$ as an exponentiation of $\hat{q}$, since
$\hat{q}$ cannot be represented on the Hilbert space $\mathcal{H}_{\textrm{poly}}$
in a well-defined manner as the generator of $\hat{U}_{\mu}$ due
to the criterion $\lim_{\mu\rightarrow0}\langle\lambda|\hat{U}_{\mu}|\lambda\rangle\neq\langle\lambda|\hat{U}_{\mu=0}|\lambda\rangle$.
In this case the basic operators $\hat{U}_{\mu},\hat{V}_{\lambda}$
act on the basis vectors $|p\rangle$ as
\begin{align}
\hat{U}_{\mu}|p\rangle= & |p-\mu\rangle,\label{eq:U-lamb-prime}\\
\hat{V}_{\lambda}|p\rangle= & e^{\frac{i}{\hbar}\lambda p}|p\rangle.\label{eq:V-lamb-prime}
\end{align}
Here, one can see from \eqref{eq:U-lamb-prime} that by starting
from a certain $p=p_{0}$, the states are again restricted to a (one-dimensional)
lattice in $p$ space where the wave functions $\Psi(p)=\langle p|\Psi\rangle$
have nonvanishing values only on the lattice points $\{p_{n}|p_{n}=p_{0}+n\mu,\,n\in\mathbb{Z}\}$
for that $p_{0}$; here also we set $p_0=0$. So we see that $p$ is discrete in this polarization
and write the basis $|p_{n}\rangle$ as a countable one, labeled with
$n$ whereas the value of position is restricted to $-\frac{\pi\hbar}{\mu}\leq q< \frac{\pi\hbar}{\mu}$. We call this polarization
the $p$ polarization, in which $p$ is discrete.
The Hilbert space ${\mathcal H}_{p_0}$ now is a superselected sector of $\mathcal{H}_{\textrm{poly}}$, such that $\mathcal{H}_{\textrm{poly}}= \bigoplus_{0\leq p_0< \mu} {\mathcal H}_{p_0}$. 

At this point a remark on notation is in order. In the rest of this work, whenever we use the $q$ polarization, we adopt the notation $|q_n\rangle$ for the discrete position basis and $|p\rangle$ for the continuous ``momentum'' basis. On the other hand, for the case of the $p$ polarization, the discrete momentum basis is written as $|p_n\rangle$ and the continuous position basis as $|q\rangle$. Additionally we emphasize that from now on we are going to work in the separable superselected Hilbert spaces ${\mathcal H}_{q_0}$ or ${\mathcal H}_{p_0}$, and not in the full polymer Hilbert space.

Let us first consider the $q$ polarization in which
\begin{align}
\hat{q}|q_{n}\rangle= & q_{n}|q_{n}\rangle,\label{eq:q-descrt-polarz-1}\\
\hat{V}_{\lambda}|q_{n}\rangle= & |q_{n}-\lambda\rangle,\label{eq:q-descrt-polarz-2}\\
\hat{q} |p\rangle = & \frac{\hbar}{i}\partial_p |p\rangle,\\
\hat{V}_{\lambda}|p\rangle= & e^{\frac{i}{\hbar}\lambda p}|p\rangle.\label{eq:q-descrt-polarz-3}
\end{align}
Notice that $\langle q_{n}|p\rangle=\sqrt{\frac{\lambda}{2\pi\hbar}}e^{\nicefrac{-iq_{n}p}{\hbar}},\,\int_{-\frac{\pi\hbar}{\lambda}}^{\frac{\pi\hbar}{\lambda}}dp|p\rangle\langle p|=1,$ and $\sum_{n\in \mathbb{Z}}|q_{n}\rangle\langle q_{n}|=1$ (see Appendix \ref{sec-app:act-U-on-q}).

\begin{figure}
\centering{}\includegraphics{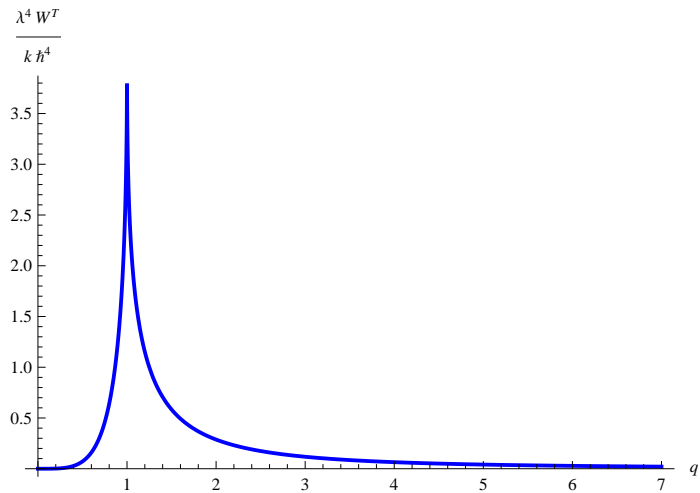}\protect\caption{The form of potential after polymerization with $\lambda=1$ in $q$ polarization. Note
that for simplicity we have plotted $\frac{W^{T}}{(k\hbar^{4}/\lambda^{4})}=\left(\sqrt{|q-1|}-\sqrt{|q+1|}\right)^{4}$, a continuous graph, while it actually
should be a discrete one. The maximum is finite and here is equal to $4$.\label{fig:PlotV-Poly}}
\end{figure}

Now, classically the Euclidean action can be written as
\begin{equation}
S_{E}=\int_{\tau_{i}}^{\tau_{f}}d\tau\left(p\left(\frac{dq}{d\tau}\right)-H_{E}(q,p)\right)
\end{equation}
with
\begin{equation}
H_{E}=\frac{p^{2}}{2m}-W(q).\label{eq:Euclid-H-clas}
\end{equation}
Using (\ref{eq:q-descrt-polarz-2}), the kinetic term in this Hamiltonian
can be represented as (see Appendix \ref{sec-app:act-q2-on-q})
\begin{equation}
p^{2}\rightarrow\widehat{p_{\lambda}^{2}}=\frac{\hbar^{2}}{\lambda^{2}}\left(2-\hat{V}_{\lambda}-\hat{V}_{-\lambda}\right).
\end{equation}
The potential $\frac{1}{q^{2}}$ in this case can be represented using
a regularization following Thiemann \citep{Thiemann1998}
\begin{align}
\frac{1}{\sqrt{|q|}}= & \frac{2}{i\lambda}V_{-\lambda}\left\{ \sqrt{|q|},V_{\lambda}\right\} \nonumber \\
= & \frac{V_{-\lambda}}{i\lambda}\left\{ \sqrt{|q|},V_{\lambda}\right\} +\left\{ \sqrt{|q|},V_{\lambda}\right\} \frac{V_{-\lambda}}{i\lambda},
\end{align}
where in the second line, we have chosen a specific symmetrization. It is obvious that other types of orderings are
also possible. The full Euclidean Hamiltonian (\ref{eq:Euclid-H-clas})
can be represented as (using the Dirac prescription $\{\cdot,\cdot\}\rightarrow-\nicefrac{i}{\hbar}[\cdot,\cdot]$;
see Appendix \ref{sec-app:rep-HE-q,V})
\begin{equation}
\hat{H}_{E}=\frac{\hbar^{2}}{2m\lambda^{2}}\left(2-\hat{V}_{\lambda}-\hat{V}_{-\lambda}\right)-k\left(\frac{\hat{V}_{-\lambda}}{\lambda}\hbar\left[\widehat{\sqrt{|q|}},\hat{V}_{\lambda}\right]+\left[\widehat{\sqrt{|q|}},\hat{V}_{\lambda}\right]\hbar\frac{\hat{V}_{-\lambda}}{\lambda}\right)^{4}.\label{eq:Euclid-H-Qreps}
\end{equation}
One can act the above Hamiltonian on basis vectors $|q_{n}\rangle$
to get (Appendix \ref{sec-app:rep-HE-q,V})
\begin{equation}
\hat{H}_{E}|q_n\rangle=\frac{\hbar^{2}}{2m\lambda^{2}}\left(2|q_{n}\rangle-|q_{n}-\lambda\rangle-|q_{n}+\lambda\rangle\right)-\frac{k\hbar^{4}}{\lambda^{4}}\left(\sqrt{|q_{n}-\lambda|}-\sqrt{|q_{n}+\lambda|}\right)^{4}|q_{n}\rangle.\label{eq:pot-Thiem-discr-q}
\end{equation}
Now the potential is not singular anymore at $q=0$
as can be seen from (\ref{eq:pot-Thiem-discr-q}) above, and it is illustrated in Fig. \ref{fig:PlotV-Poly}.
Also if the energy of the system is smaller than the peak of the potential and the particle's initial position is such that $0\leq q_0 < q(W_{\textrm{max}})$ where $q(W_{\textrm{max}})$ is the position corresponding to the potential peak,
the semiclassical issue mentioned above is solved: since $q$ remains finite
at all times, then action if functionally differentiable (boundary terms vanish) and $S[q_{cl}]<\infty$.
Therefore there is no need to add a counterterm to the action to
be able to do a saddle point approximation. However, in a more general
case, when the energy of the system is greater than the peak of the
potential, we have
\begin{equation}
q_{cl}\big|_{t\rightarrow\infty}\rightarrow\infty
\end{equation}
and due to (\ref{eq:SE-Leib}), we
will  still have the problem that  the action
will not necessarily be functionally differentiable and that even so,
$S_{E}[q_{cl}]$ will not be finite.

This analysis shows that this kind of polymerization, which does not
bound the position but discretizes it, removes
the singularity of the potential at the origin. However it does not solve
the problems with the saddle point approximation, and we still need
to add a Hamilton-Jacobi counterterm to get a well-defined action
for this kind of approximation. 

One solution to both of the above problems (both functional differentiability of the action and/or its finiteness on  classical solutions) comes from modifying
the potential at infinity such that $q_{t\rightarrow\infty}<\infty$
and consequently $\delta q=0$ at the (time) boundary. This is the
subject of the next section. In the remaining part of this section,
we investigate the effects of polymerization on the Hamilton-Jacobi
counterterm in $q$ polarization. To do this, we first derive the
effective Euclidean Hamiltonian in $q$ polarization using the path integral
method. This turns out to be (see Appendix \ref{sec-app:eff-act})
\begin{equation}
H_{\textrm{eff}}=\frac{2\hbar^{2}}{m\lambda^{2}}\sin^{2}\left(\frac{\lambda p}{2\hbar}\right)-W^{T}=\frac{2\hbar^{2}}{m\lambda^{2}}\sin^{2}\left(\frac{\lambda p}{2\hbar}\right)-\frac{k\hbar^{4}}{\lambda^{4}}\left(\sqrt{|q-\lambda|}-\sqrt{|q+\lambda|}\right)^{4},\label{eq:H-eff-Thiem}
\end{equation}
where $W^{T}$ is the continuous counterpart of the potential derived
using Thiemann's regularization. A note about some subtleties is in
order here: we could not directly argue that the form of the effective
Hamiltonian will be (\ref{eq:H-eff-Thiem}) based on the form of the
potential in (\ref{eq:pot-Thiem-discr-q}). This is so because if
one computes the action of the full quantum Hamiltonian on $|q_{n}\rangle$
states, one gets the above effective potential but with discrete $q_{n}$,
and also the kinetic term will not be $\frac{2\hbar^{2}}{m\lambda^{2}}\sin^{2}\left(\frac{\lambda p}{2\hbar}\right)$
anymore since this is the form of the kinetic term when it acts on
$|p\rangle$ states and not the $|q_{n}\rangle$ ones. On the other hand,
if we use the $|p\rangle$ basis for the action of the full quantum Hamiltonian,
we would not get the effective potential in (\ref{eq:H-eff-Thiem}).
In the first case (acting $\hat{H}$ on $|q_{n}\rangle$), one will
get an additional $\int dp$ in the kinetic term (and also the potential
will be discrete), while in the second case (acting $\hat{H}$ on $|p\rangle$),
there will be a $\sum_{n}$ in the potential term. The way to overcome
these problems and get the effective action (\ref{eq:H-eff-Thiem})
is to use the path integral formulation and not something like $H_{\textrm{eff}}=\langle q_{n}|H|q_{n}\rangle$.
What makes the path integral method useful is that one can
bring the extra $\sum_{n}$ out of the exponential of the action
and turn it into an integral that appears in the measure of the path
integral (see Appendix \ref{sec-app:eff-act} for more details). 

To find the explicit form of the polymer $G_{\textrm{poly}}$ for this effective Hamiltonian,
we need to solve (\ref{eq:HJ-eq-gen}). Considering the above Hamiltonian
and noting that $p=\frac{\partial G_{\textrm{poly}}}{\partial q}$, Eq. (\ref{eq:HJ-eq-gen})
becomes
\begin{equation}
\frac{2\hbar^{2}}{m\lambda^{2}}\sin^{2}\left(\frac{\lambda}{2\hbar}\frac{\partial G_{\textrm{poly}}}{\partial q}\right)-\frac{k\hbar^{4}}{\lambda^{4}}\left(\sqrt{|q-\lambda|}-\sqrt{|q+\lambda|}\right)^{4}+\frac{\partial G_{\textrm{poly}}}{\partial t}=0.\label{eq:HJ-inv-pot-poly}
\end{equation}
Again since the Hamiltonian does not depend explicitly on $t$, Eq. (\ref{eq:HJ-eq-gen})
implies that the derivative $\nicefrac{\partial G_{\textrm{poly}}}{\partial t}$ must
be a constant $H=E$, and an additive separation of variables in $G_{\textrm{poly}}$
can be performed
\begin{equation}
G_{\textrm{poly}}(q,E,t)=\Omega(q,E)-Et.\label{eq:G-Omega-E-poly}
\end{equation}
This turns (\ref{eq:HJ-inv-pot-poly}) into 
\begin{equation}
\frac{2\hbar^{2}}{m\lambda^{2}}\sin^{2}\left(\frac{\lambda}{2\hbar}\frac{\partial\Omega}{\partial q}\right)-\frac{k\hbar^{4}}{\lambda^{4}}\left(\sqrt{|q-\lambda|}-\sqrt{|q+\lambda|}\right)^{4}-E=0\label{eq:HJ-no-t-deriv-poly}
\end{equation}
or
\begin{equation}
\frac{\partial\Omega}{\partial q}=\frac{2}{\lambda}\sin^{-1}\left(\sqrt{\frac{m\lambda^{2}}{2\hbar^{2}}E+\frac{mk\hbar^{2}}{2\lambda^{2}}\left(\sqrt{|q-\lambda|}-\sqrt{|q+\lambda|}\right)^{4}}\right).\label{eq:dOmega-dq-poly}
\end{equation}
As clearly seen from this, finding $\Omega$ in this case is
much more involved due to the presence of $\sin^{-1}\sqrt{\cdots}$.
A way around this difficulty is to use a perturbative expansion of the right hand
side of the above, around $\nicefrac{\lambda}{q}=0$.
This yields
\begin{equation}
\frac{\partial\Omega}{\partial q}=\frac{\sqrt{2}\sqrt{m}\sqrt{Eq^{2}+k}}{q}+\frac{\lambda^{2}}{12\hbar^{2}q^{3}}\left(\sqrt{2}m^{\nicefrac{3}{2}}\left(Eq^{2}+k\right)^{\nicefrac{3}{2}}+\frac{3k\hbar^{2}\sqrt{2}\sqrt{m}}{\sqrt{Eq^{2}+k}}\right)+\mathcal{O}\left(\frac{\lambda^{4}}{\hbar^{4}}\right)\label{dOmdq}
 \end{equation}
The leading continuum order in (\ref{dOmdq}), in which $\lambda=0$,
matches exactly the corresponding classical counterpart in (\ref{eq:w-in-HJ}).
Integrating (\ref{dOmdq}) and then substituting the result into (\ref{eq:G-Omega-E-poly})
yields
\begin{align}
G_{\textrm{poly}}= & -Et+\sqrt{2}\sqrt{m}\sqrt{Eq^{2}+k}+\sqrt{2}\sqrt{mk}\tan^{-1}\left(\sqrt{\frac{k}{Eq^{2}+k}}\right)\\
 & +\frac{\lambda^{2}}{\hbar^{2}}\left[\frac{\sqrt{2}\sqrt{m}\sqrt{Eq^{2}+k}}{24q^{2}}\left(m\left(2Eq^{2}-k\right)-3\hbar^{2}\right)+\frac{\sqrt{2}\sqrt{m}}{8\sqrt{k}}E(+mk+\hbar^{2})\tan^{-1}\left(\sqrt{\frac{k}{Eq^{2}+k}}\right)\right]+\mathcal{O}\left(\frac{\lambda^{4}}{\hbar^{4}}\right).\label{eq:H-J-poly-fin-E-explicit}
\end{align}
Again we see that the purely classical term with $\lambda=0$ matches
exactly to the classical Hamilton's principal function (\ref{eq:G-explic})
while there are also several types of corrections due to polymer quantization. 

Since the classical part of the polymerized $G_{\textrm{poly}}$ in (\ref{eq:H-J-poly-fin-E-explicit})
is exactly the same as the classical nonpolymerized case, and since
we have seen that with purely classical $G$, the action is functionally differentiable and its value on classical solutions is finite, we conclude that the polymerized improved
action $\Gamma_{\textrm{poly}}[q_{cl}]$ with the above counterterm
has also the same nice properties and thus makes it possible to proceed
with the saddle point approximation if desired. But in addition to
that, the counterterm (\ref{eq:H-J-poly-fin-E-explicit}) has additional
terms proportional to the powers of ``quantum lattice parameter''
$\lambda$. So, it is reasonable to expect that this, together with
the change of the form of the bulk action due to polymer quantization,
will change the thermodynamical properties of the system in case of,
e.g., a black hole, and therefore it is interesting to see what are
the implications of such semiclassical polymer modifications. Next we consider the other polymer polarization.

\section{The polymer model with discrete momentum \label{sec:eff-poly-without-HJ}}

In this section we consider the $p$ polarization in which
\begin{align}
\hat{U}_{\mu}|p_{n}\rangle= & |p_{n}-\mu\rangle,\label{eq:p-descrt-polarz-1}\\
\hat{p}|p_{n}\rangle= & p_{n}|p_{n}\rangle,\label{eq:p-descrt-polarz-2}\\
\hat{U}_{\mu}|q\rangle= & e^{\frac{i}{\hbar}\mu q}|q\rangle,\label{eq:p-descrt-polarz-3}\\
\hat{p} |q\rangle = & -\frac{\hbar}{i}\partial_q |q\rangle.
\end{align}
%Here also, the third equation can be taken as a first principle or can be derived using the first two (see appendix \ref{sec-app:act-U-on-q}).}
The representation of the kinetic term in (\ref{eq:Euclid-H-clas})
in this polarization is very simple; in fact it is just $\frac{\hat{p}^{2}}{2m}$.
The problem here is how to represent the potential. It is not clear
how Thiemann's regularization can be used in this case. The reason
is that generally this regularization is used to represent a variable that is discrete and not bounded. In the present case though we have to define a replacement for the inverse of $\hat{q}$; a problematic task since even $\hat{q}$ is not well defined on ${\cal{H}}_{p_0}$but only $\hat{U}_{\mu}$. Even more, we may consider finding functions $F(U_{\mu})$ and $G(p)$ such that classically
\begin{equation}\label{FUGp}
\frac{1}{q^{n}}=\left\{ F\left(U_{\mu}\right),G(p)\right\} ^{m},\,\,\,\,\,\,\,\,\,\,\,\,\,\,n,m>0,
\end{equation}
so that $F(U_{\mu})$ and $G(p)$  admit a simple representation on Hilbert space. The first part, i.e., finding classical functions
$F(U_{\mu})$ and $G(p)$ fulfilling (\ref{FUGp}) may not be
very hard, and several options may be available such as
\begin{equation}
\frac{2}{\sqrt{\mu}}\left\{ \sqrt{\big|-i\ln\left(U_{\mu}\right)\big|},p\right\} =\frac{2}{\sqrt{\mu}}\left\{ \sqrt{\mu}\sqrt{\big|q\big|},p\right\} =\frac{1}{\sqrt{|q|}}.
\end{equation}
However, the second part, the ability to represent the functions $F(U_{\mu})$
and $G(p)$ on Hilbert space, is the hard part as can be seen from
the above example. Presently we have not found satisfactory functions
that can be represented on $\mathcal{H}_{\textrm{poly}}$ for which Thiemann's regularization
can be done. In spite of this difficulty it is still possible to introduce a formal inverse squared position operator in this polarization and hence its semiclassical approximation.

We saw in the previous subsection that a solution to both issues of functional differentiability and finiteness of the action is likely to come from bounding $q$ to finite values, and this may result from modifying the potential term in $H_{E}$. Considering this, 
%and despite the problem that we presently can not represent the potential
%$W$ in this $(\hat{U}_{\mu},\hat{p})$ polarization, 
we can argue that since in this polarization,
$q$ is bounded, a representation $\hat{W}(\hat{U}_{\mu})$
is expected to lead to a bounded potential and thus would eliminate the need
to add a Hamilton-Jacobi boundary counterterm to the action to enable
one to make a well-defined saddle point approximation. 
%But in any case, this remains to be seen in a concrete mathematical way after a suitable representation of $W$ is found. 
Now we provide a scheme that shows how the polymer quantization will change
the action in such a way that no counterterm is needed to have a
well-defined saddle point approximation.

Our proposed scheme for the effective potential is based on the observation
that $q^{2}$ in this polarization gets replaced by an operator that is represented as (see Appendix
(\ref{sec-app:act-q2-on-q}))
\begin{equation}
q^{2}\rightarrow\widehat{q_{\mu}^{2}}=\frac{\hbar^2}{\mu^2}\left(2-\hat{U}_{\mu}-\hat{U}_{-\mu}\right).
\end{equation}
Its action on a $|q\rangle$ basis is
\begin{align}
\frac{\hbar^2}{\mu^{2}}\left(2-\hat{U}_{\mu}-\hat{U}_{-\mu}\right)|q\rangle= & \frac{4\hbar^2}{\mu^{2}}\sin^{2}\left(\frac{\mu q}{2\hbar}\right)|q\rangle.
\end{align}This is computed using (\ref{eq:p-descrt-polarz-3}) (see Appendix
(\ref{sec-app:act-q2-on-q})). Based on this, our proposed replacement of the inverse squared 
operator is 
\begin{equation}
\widehat{\frac{1}{q_{\mu}^{2}}}|q\rangle=\frac{\mu^{2}}{4\hbar^{2}}\csc^{2}\left(\frac{\mu q}{2\hbar}\right)|q\rangle.\label{eq:rep-1-over-q2}
\end{equation}%where $\widehat{\frac{1}{q_{\mu}^{2}}}$ should be considered as a symbolic expression used for the representation of the classical function $\frac{1}{q^{2}}$ on the Hilbert space. 
Using path integral formulation,
the effective action turns out to be (Appendix \ref{sec-app:eff-act})
\begin{equation}
S_{\textrm{eff}}=\int dt\left[p\dot{q}-\left(\frac{p^{2}}{2m}-\frac{\mu^{2}k}{4\hbar^{2}}\csc^{2}\left(\frac{\mu q}{2\hbar}\right)\right)\right].\label{eq:act-eff-PI}
\end{equation}Eq. (\ref{eq:act-eff-PI}) suggests that the effective form
of the classical potential $W(q)=\frac{k}{q^{2}}$ in this scheme
has an effective form 
\begin{equation}
W^{h}=\frac{\mu^{2}k}{4\hbar^{2}}\csc^{2}\left(\frac{\mu q}{2\hbar}\right).\label{eq:heur-potential}
\end{equation}Note that here also there are subtleties similar to those explained
in the previous section related to the important role of the path
integral method to derive (\ref{eq:act-eff-PI}). For example, acting
quantum Hamiltonian on $|q\rangle$ yields the above effective potential
while the kinetic term will not be $\nicefrac{p^{2}}{2m}$ anymore
since this is the form of the kinetic term when it acts on $|p\rangle$
not on $|q\rangle$ . Furthermore, since in this polarization our momentum
states are actually discrete, i.e., $|p_{n}\rangle$, $p_n=n\mu, n\in\mathbb{Z}$, even if we act
the kinetic term on them instead of $|q\rangle$, we will get a discrete
result and not a continuous one. Thus again the path integral method saves
the day and makes it possible to get rid of the additional $\sum_{n}$
and the discreteness of the momentum in the effective action. Details
are explained in Appendix \ref{sec-app:eff-act}.

\begin{figure}
\centering{}\includegraphics{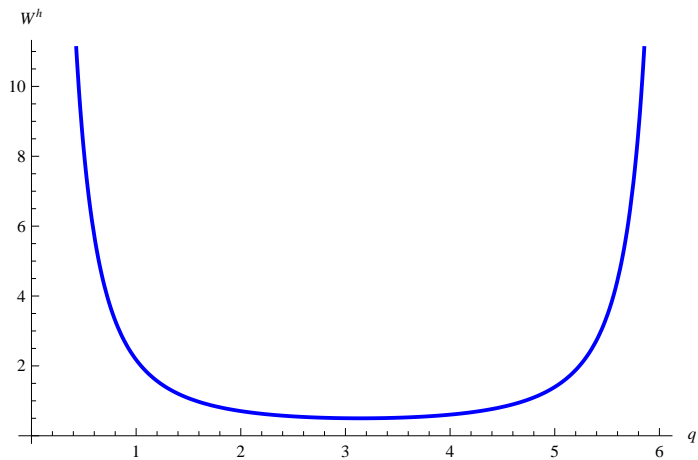}\protect\caption{The form of potential $W^h$ for our model with $\mu=1$.\label{fig:PlotV-Poly-csc}}
\end{figure}

The plot of the potential (\ref{eq:heur-potential}) is shown in Fig.
\ref{fig:PlotV-Poly-csc}. Obviously the potential is still singular
at the origin, $q=0$. But the good news is that 
the position is bounded at the time boundary, i.e., $q_{cl}(t\rightarrow\infty)<\infty$.
This means that since now $q$ is finite at the boundary, $\delta q=0$
on the boundary, and thus the action is welldefined for a saddle point
approximation without the need to add any other terms, such as the
Hamilton-Jacobi counterterm. This polarization, like the previous
one, will also most probably lead to modifications to the thermodynamics
of the system due to the modification of the bulk action due to polymer
quantization as well as the absence of any counterterm. This is particularly interesting if the system under study is a black hole.

\section{Discussion\label{sec:conclude}}

In this work we have studied the issue of ill-defined saddle point semiclassical approximation that occurs for some systems including
dilatonic black holes and some mathematically analog  mechanical models. This problem
arises due to the fact that the  action is not functionally differentiable for all the variations
of the fields, compatible with the boundary and falloff conditions  and even if so, the value of the action on classical
solutions does not remain finite for all of these conditions. This issue is rather important since one of the
main methods of deriving thermodynamical properties of such models
is through the saddle point approximation to the path integral that
can also be interpreted as the partition function of the system. The
common solution to this problem is to add a boundary counterterm
to the action which is a solution to the corresponding Hamilton-Jacobi
equation. A very interesting observation is that only with this term can one get the correct thermodynamics for certain black holes.

We undertake to seek an alternative method to attack this issue which
is through polymer quantization of the model. The effects of polymerization
may lead to two outcomes: either it will spare us altogether from
adding any additional term to the action and the action is already
well defined for saddle point approximation after polymer modifications,
or it may not remove the necessity of adding a counterterm, but it
will modify it. In both cases, it is very likely that the process
of polymer quantization will change the thermodynamics, either due
to the fact the counterterm is not necessary or because it will be
modified.

In this work we restrict our study to a simpler analog model featuring the aforementioned issues. This helps
us see the effects, problems, and results of this method more clearly
and provides interesting insights about the procedure. The analog
model we study is a single particle in an inverse square potential. 

We first show how this system is modified under different polarizations
of polymer quantization. It turns out that in the polarization where
$q$ is discrete, we still will need to add the Hamilton-Jacobi counterterm
to the action, but the advantage is that the potential can be rather
easily represented on a Hilbert space using Thiemann's regularization
\citep{Thiemann1998}. This is the case in which the Hamilton-Jacobi
counterterm is modified by polymerization. We then proceed to compute
the effective action and thus derive the effective Hamiltonian using
path integral formulation. Then we derive the associated Hamilton-Jacobi
counterterm and show that the classical terms of the polymerized
case match exactly the nonpolymerized case while there are corrections
to this counterterm that come from the polymer quantization. This
supports our claim that polymer quantization will change the thermodynamics
of the system due to the polymer modifications of the bulk action
and the Hamilton-Jacobi counterterm.

On the other hand, in the polarization where $p$ is discrete, we
argue that in general there should be no need to add a counterterm to the action since
the variable $q$ is bounded due to polymer effects; thus the action
remains functionally differentiable and finite on classical solutions in accordance with
all the allowed variations. However, since the representation of the
potential in this case is not so straightforward and could not be done
here, in order to get concrete results and calculations,
we proposed an effective form that replaces the classical potential $W=\frac{1}{q^{2}}$
based on the analysis of the polymer operator $\widehat{q^{2}_{\mu}}$. Using this, we show that the effective action is indeed finite evaluated on the effective solution and is functionally differentiable without the need
to add a counterterm. This means that due to the effects of polymer
quantization in $p$ polarization, the system is already well defined
for saddle point approximation. 

Further developments along the lines of the present work include the following. It is pretty evident that a similar analysis for the case of polymer black hole systems may lead to the change of thermodynamics due to polymer modifications
of the bulk action. It will be interesting to check whether any counterterm is required in such a case.
Also, recently, much work has been done to unveil singularity avoidance in loop quantum cosmology \citep{Ashtekar:2010gz} as well as in black holes \citep{bohmer2007loop,corichi2015loop,chiou2008phenomenological} by alluding to effective models like the ones we have studied in the present work; hence it would be interesting to explore possible physical consequences for a semiclassical approximation of these systems using the path integral framework.

\begin{acknowledgments}
The authors would like to thank Daniel Grumiller for his enlightening
comments on Ref. \citep{Grumiller2007a}. They would also like to acknowledge the partial support
of CONACyT Grant No. 237351: Implicaciones F\'{i}sicas de la Estructura
del Espaciotiempo. S.R. would like to acknowledge the support of the
 PROMEP postdoctoral fellowship (through UAM-I) and the grant from
Sistema Nacional de Investigadores of CONACyT. D.H.O.B. would like
to acknowledge the support of CONACyT Grant No. 283451.
\end{acknowledgments}

\appendix

\section{Proof of (\ref{eq:q-descrt-polarz-3}) and (\ref{eq:p-descrt-polarz-3})
\label{sec-app:act-U-on-q}}

Let us consider the polarization $(\hat{q},\hat{V}_{\lambda})$ where
eigenvalues of the operator $\hat{q}$ are discrete. We would like
to derive (\ref{eq:q-descrt-polarz-3}) from (\ref{eq:q-descrt-polarz-1})
and (\ref{eq:q-descrt-polarz-2}). Using the completeness relation 
\begin{equation}
1=\frac{\lambda}{2\pi\hbar}\int_{-\nicefrac{\pi\hbar}{\lambda}}^{\nicefrac{\pi\hbar}{\lambda}}dp|p\rangle\langle p|
\end{equation}
and the form of 
\begin{equation}
\langle q_{n}|p\rangle=\sqrt{\frac{\lambda}{2\pi\hbar}}e^{\frac{ip_{n}q}{\hbar}}
\end{equation}
one can write
\begin{align}
\hat{V}_{\lambda}|p\rangle= & \sum_{n=-\infty}^{\infty}\hat{V}_{\lambda}|q_{n}\rangle\langle q_{n}|p\rangle\nonumber \\
= & \frac{\lambda}{2\pi}\int_{-\nicefrac{\pi}{\lambda}}^{\nicefrac{\pi}{\lambda}}dp^{\prime}\sum_{n=-\infty}^{\infty}|p^{\prime}\rangle\langle p^{\prime}|q_{n}-\lambda\rangle\sqrt{\frac{\lambda}{2\pi\hbar}}e^{\frac{i}{\hbar}pq_{n}}\nonumber \\
= & \frac{\lambda}{2\pi}\int_{-\nicefrac{\pi}{\lambda}}^{\nicefrac{\pi}{\lambda}}dp^{\prime}\sum_{n=-\infty}^{\infty}|p^{\prime}\rangle e^{-\frac{i}{\hbar}p^{\prime}q_{n}}e^{\frac{i}{\hbar}p^{\prime}\lambda}e^{\frac{i}{\hbar}pq_{n}}\nonumber \\
= & \lambda\int_{-\nicefrac{\pi}{\lambda}}^{\nicefrac{\pi}{\lambda}}dp^{\prime}\lim_{j\rightarrow\infty}\underbrace{\left[\frac{1}{2\pi}\sum_{n=-j}^{j}e^{\frac{i}{\hbar}q_{n}(p-p^{\prime})}\right]}_{\delta_{n}(p-p^{\prime})}e^{\frac{i}{\hbar}p^{\prime}\lambda}|p^{\prime}\rangle,
\end{align}
where $\delta_{n}(p-p^{\prime})$ is a Dirac delta sequence and the
limit will make it into a formal Dirac delta series representation
$\delta(p-p^{\prime})$. Using the above and the fact that for a function
$f(p)$ with a period $\nicefrac{2\pi}{\lambda}$ one can write 
\begin{equation}
\int_{-\nicefrac{\pi}{\lambda}}^{\nicefrac{\pi}{\lambda}}dp^{\prime}\delta(p-p^{\prime})f(p^{\prime})=\frac{1}{\lambda}f(p),
\end{equation}
we finally arrive at
\begin{align}
\hat{V}_{\lambda}|p\rangle= & e^{\frac{i}{\hbar}p\lambda}|p\rangle.
\end{align}
Using the same lines of argument but now in the polarization $(\hat{U}_{\mu},\hat{p})$
where eigenvalues of the operator $\hat{p}$ are discrete, one can
derive (\ref{eq:p-descrt-polarz-3}) from (\ref{eq:p-descrt-polarz-1})
and (\ref{eq:p-descrt-polarz-2}).

\section{Action of $\widehat{q_{\mu}^{2}}|q\rangle$ and $\widehat{p_{\lambda}^{2}}|p\rangle$\label{sec-app:act-q2-on-q}}

Let us choose one of the polarizations, say the one in which $p$
is discrete. Since as mentioned before, the $\hat{q}$ operator does
not exist in this polarization, we first need to construct an analog
of the $q^{2}$ operator that we call $\widehat{q_{\mu}^{2}}$ and
then find its action on the desired basis. To do this, we note that
since classically $U_{\mu}=e^{\nicefrac{i\mu q}{\hbar}}$, then one
can write
\begin{equation}
e^{\frac{i\mu q}{\hbar}}+e^{-\frac{i\mu q}{\hbar}}\approx2-\frac{\mu^{2}q^{2}}{\hbar^{2}},\hspace*{1cm}q\ll\frac{\hbar}{\mu}.
\end{equation}
Using this, one can isolate $q^{2}$ and represent its singular counterpart $\widehat{q_{\mu}^{2}}$ (singular here means the limit $\mu\rightarrow0$
does not exist) on the Hilbert space as 
\begin{equation}
\widehat{q_{\mu}^{2}}=\frac{\hbar^{2}}{\mu^{2}}\left(2-\hat{U}_{\mu}-\hat{U}_{-\mu}\right),\label{eq:q2mu-op-def}
\end{equation}
and using the results of Appendix \ref{sec-app:act-U-on-q} or equivalently
using (\ref{eq:p-descrt-polarz-3}), one can write
\begin{align}
\frac{\hbar^{2}}{\mu^{2}}\left(2-\hat{U}_{\mu}-\hat{U}_{-\mu}\right)|q\rangle= & \frac{\hbar^{2}}{\mu^{2}}\left(2-e^{\frac{i}{\hbar}\mu q}-e^{-\frac{i}{\hbar}\mu q}\right)|q\rangle\\
= & \frac{\hbar^{2}}{\mu^{2}}\left(2-2\cos\left(\frac{\mu q}{\hbar}\right)\right)|q\rangle\\
= & \frac{4\hbar^{2}}{\mu^{2}}\left(\sin^{2}\left(\frac{\mu q}{2\hbar}\right)\right)|q\rangle.
\end{align}
So we conclude that 
\begin{equation}
\widehat{q_{\mu}^{2}}|q\rangle=\frac{4\hbar^{2}}{\mu^{2}}\left(\sin^{2}\left(\frac{\mu q}{2\hbar}\right)\right)|q\rangle.\label{eq:q2mu-op-act}
\end{equation}
On the other hand, if we choose to work in the polarizations in which
$q$ is discrete, the operator $\widehat{p_{\lambda}^{2}}$ can be
defined and represented using the same method, i.e.,
\begin{equation}
\widehat{p_{\lambda}^{2}}=\frac{\hbar^{2}}{\lambda^{2}}\left(2-\hat{V}_{\lambda}-\hat{V}_{-\lambda}\right).\label{eq:p2lambda-op-def}
\end{equation}
Its action on the $|p\rangle$ basis can then be computed in the same
manner as above for $\widehat{q_{\mu}^{2}}$, and by applying the results
of Appendix \ref{sec-app:act-U-on-q} or equivalently Eq. (\ref{eq:q-descrt-polarz-3})
one gets
\begin{equation}
\widehat{p_{\lambda}^{2}}|p\rangle=\frac{4\hbar^{2}}{\lambda^{2}}\left(\sin^{2}\left(\frac{\lambda p}{2\hbar}\right)\right)|p\rangle.\label{eq:p2lambda-op-act}
\end{equation}

\section{Effective polymerized actions\label{sec-app:eff-act}}

In this section we show how to obtain the effective action of the
models described in Secs. \ref{sec:eff-poly-with-HJ} and \ref{sec:eff-poly-without-HJ}.
The calculations are done in Newtonian form but changing to the
Euclidean form is straightforward. In steps that are different for each
polarization, we will make comments and make clear what equation corresponds
to which polarization. The transition amplitude in each case 
is given by 
\begin{align}
\langle q_{f},t_{f}\mid q_{i},t_{i}\rangle= & \lim_{N\rightarrow\infty}\left(\prod_{n=1}^{N-1}\sum_{n}\right)\left(\prod_{n=1}^{N}\langle q_{n},t_{n}\mid q_{n-1},t_{n-1}\rangle\right),\,\,\,\,\,\,\,\,\,\,\,\,\,\,\,\,\,\,\,\,\,\,\,\,\,\,\,(\textrm{Discrete }q),\label{eq:full-trans-amp-discrt-q}\\
\langle q_{f},t_{f}\mid q_{i},t_{i}\rangle= & \lim_{N\rightarrow\infty}\left(\prod_{n=1}^{N-1}\int_{-\frac{\pi\hbar}{\mu}}^{\frac{\pi\hbar}{\mu}}dq_{n}\right)\left(\prod_{n=1}^{N}\langle q_{n},t_{n}\mid q_{n-1},t_{n-1}\rangle\right),\,\,\,\,\,\,\,\,\,\,\,\,\,\,\,\,\,\,\,\,\,\,\,\,\,\,\,(\textrm{Discrete }p).\label{eq:full-trans-amp-discrt-p}
\end{align}
where in the first equation, we have a sum
over polymer lattice $q$ points due to discreteness of $q$ and in
the second equation, the limits of the integral reflect the bounds on
the continuous variable $q$. Note that in the first equation, the
subscript is due to the genuine polymer lattice discreteness of $q$, while
in the second one, it is due to partitioning of the path integral. The
amplitude above is then divided into partitions for which we have
\begin{align}
\langle q_{n},t_{n}\mid q_{n-1},t_{n-1}\rangle= & \langle q_{n}\mid e^{-\frac{i\epsilon}{\hbar}\hat{H}_{\textrm{poly}}}\mid q_{n-1}\rangle\nonumber \\
\approx & \langle q_{n}\mid1-\frac{i\epsilon}{\hbar}\hat{H}_{\textrm{poly}}\mid q_{n-1}\rangle\nonumber \\
= & \langle q_{n}\mid q_{n-1}\rangle-\frac{i\epsilon}{\hbar}\langle q_{n}\mid\frac{\hbar^{2}}{2m}\left(\frac{2-\hat{V}_{\lambda}-\hat{V}_{-\lambda}}{\lambda^{2}}\right)\mid q_{n-1}\rangle-\frac{i\epsilon}{\hbar}\langle q_{n}\mid\hat{W}\mid q_{n-1}\rangle,\label{eq:interval-poly-1}
\end{align}
where $\hat{H}_{\textrm{poly}}$ is the polymer quantum Hamiltonian
operator and we have used (\ref{eq:p2lambda-op-def}) for the representation
of the kinetic term. $\hat{W}$ is the potential operator that can
be represented by Thiemann's regularization as in (\ref{eq:Euclid-H-Qreps})
in case $q$ is discrete or can be the potential proposed in Sec. \ref{sec:eff-poly-without-HJ}
when $p$ is discrete. We then insert one of the following identities:
\begin{align}
1= & \sum_{n}|p_{n}\rangle\langle p_{n}|,\,\,\,\,\,\,\,\,\,\,\,\,\,\,\,\,\,\,\,\,\,\,\,\,\,\,\,(\textrm{Discrete }p),\\
1= & \frac{\lambda}{2\pi\hbar}\int_{-\frac{\pi\hbar}{\lambda}}^{\frac{\pi\hbar}{\lambda}}dp_{n}|p_{n}\rangle\langle p_{n}|,\,\,\,\,\,\,\,\,\,\,\,\,\,\,\,\,\,\,\,\,\,\,\,\,\,\,\,(\textrm{Discrete }q),
\end{align}
in front of the kinetic and potential terms in (\ref{eq:interval-poly-1}),
corresponding to the polarization we are working in, to obtain either
\begin{equation}
\langle q_{n},t_{n}\mid q_{n-1},t_{n-1}\rangle\approx\frac{\lambda}{2\pi\hbar}\int_{-\frac{\pi\hbar}{\lambda}}^{\frac{\pi\hbar}{\lambda}}dp_{n}\langle q_{n}\mid p_{n}\rangle\langle p_{n}\mid q_{n-1}\rangle\left[1-\frac{i\epsilon}{\hbar}\left(\frac{2\hbar^{2}}{m\lambda^{2}}\sin^{2}\left(\frac{\lambda p}{2\hbar}\right)+W_{n}^{T}\right)\right]\,\,\,\,\,\,\,\,\,(\textrm{Discrete }q)\label{eq:full-amp-discrt-q-0}
\end{equation}
or
\begin{align}
\langle q_{n},t_{n}\mid q_{n-1},t_{n-1}\rangle\approx & \sum_{n}\langle q_{n}\mid p_{n}\rangle\langle p_{n}\mid q_{n-1}\rangle\left[1-\frac{i\epsilon}{\hbar}\left(\frac{p_{n}^{2}}{2m}+W_{n}^{h}\right)\right]\,\,\,\,\,\,\,\,\,\,\,\,\,\,\,\,\,\,\,\,\,\,\,\,\,\,\,(\textrm{Discrete }p).\label{eq:full-amp-discrt-p-0}
\end{align}
Here
\begin{align}
W_{n}^{T}= & \frac{k\hbar^{4}}{\lambda^{4}}\left(\sqrt{|q_{n}-\lambda|}-\sqrt{|q_{n}+\lambda|}\right)^{4},\label{eq:WT-n}\\
W_{n}^{h}= & \frac{\mu^{2}k}{4\hbar^{2}}\csc^{2}\left(\frac{\mu q_{n}}{2\hbar}\right)\label{eq:Wh-n}
\end{align}
are Thiemann-regularized and heuristic polymer potentials, respectively.
Note that again there are two types of subscripts associated with two
types of discreteness. One is related to partitioning the full transition
amplitude [as in (\ref{eq:Wh-n})], and the other is related to the
genuine polymer lattice discretization [as in (\ref{eq:WT-n})]. Using
\begin{align}
\langle p\mid q_{n}\rangle= & \sqrt{\frac{\lambda}{2\pi\hbar}}e^{-\frac{ipq_{n}}{\hbar}}\,\,\,\,\,\,\,\,\,\,\,\,\,\,\,\,\,\,\,\,\,\,\,\,\,\,\,(\textrm{Discrete }q),\\
\langle p_{n}\mid q\rangle= & \sqrt{\frac{\mu}{2\pi\hbar}}e^{-\frac{ip_{n}q}{\hbar}}\,\,\,\,\,\,\,\,\,\,\,\,\,\,\,\,\,\,\,\,\,\,\,\,\,\,\,(\textrm{Discrete }p),
\end{align}
and also the full expansion of the exponential, one gets for the above
partition transition amplitudes 
\begin{align}
\langle q_{n},t_{n}\mid q_{n-1},t_{n-1}\rangle= & \left(\frac{\lambda}{2\pi\hbar}\right)^{2}\int_{-\frac{\pi\hbar}{\lambda}}^{\frac{\pi\hbar}{\lambda}}dp_{n}e^{\frac{i\epsilon}{\hbar}\left[p_{n}\frac{(q_{n}-q_{n-1})}{\epsilon}-\left(\frac{2\hbar^{2}}{m\lambda^{2}}\sin^{2}\left(\frac{\lambda p}{2\hbar}\right)+W_{n}^{T}\right)\right]}\,\,\,\,\,\,\,\,\,\,\,\,\,\,\,\,\,\,\,\,\,\,\,\,\,\,\,(\textrm{Discrete }q),\\
\langle q_{n},t_{n}\mid q_{n-1},t_{n-1}\rangle= & \frac{\mu}{2\pi\hbar}\sum_{n}e^{\frac{i\epsilon}{\hbar}\left[p_{n}\frac{q_{n}-q_{n-1}}{\epsilon}-\left(\frac{p_{n}^{2}}{2m}+W_{n}^{h}\right)\right]}\,\,\,\,\,\,\,\,\,\,\,\,\,\,\,\,\,\,\,\,\,\,\,\,\,\,\,(\textrm{Discrete }p).
\end{align}
Using these in corresponding full transition amplitudes (\ref{eq:full-trans-amp-discrt-q})
and (\ref{eq:full-trans-amp-discrt-p}) yields 
\begin{align}
\langle q_{f},t_{f}|q_{i},t_{i}\rangle= & \lim_{N\rightarrow\infty,\epsilon\rightarrow0}\left(\prod_{n=1}^{N-1}\sum_{n}\right)\left(\left(\frac{\lambda}{2\pi\hbar}\right)^{2}\prod_{n=1}^{N}\int_{-\frac{\pi\hbar}{\lambda}}^{\frac{\pi\hbar}{\lambda}}dp_{n}\right)e^{\sum_{n=1}^{N}\frac{i\epsilon}{\hbar}\left[p_{n}\frac{(q_{n}-q_{n-1})}{\epsilon}-\left(\frac{2\hbar^{2}}{m\lambda^{2}}\sin^{2}\left(\frac{\lambda p}{2\hbar}\right)+W_{n}^{T}\right)\right]}\,\,\,\,(\textrm{Discrete }q),\label{eq:full-amp-discrt-q-1}\\
\langle q_{f},t_{f}|q_{i},t_{i}\rangle= & \lim_{N\rightarrow\infty,\epsilon\rightarrow0}\left(\prod_{n=1}^{N-1}\int_{-\frac{\pi\hbar}{\mu}}^{\frac{\pi\hbar}{\mu}}dq_{n}\right)\left(\frac{\mu}{2\pi\hbar}\prod_{n=1}^{N}\sum_{n}\right)e^{\sum_{n=1}^{N}\frac{i\epsilon}{\hbar}\left[p_{n}\frac{q_{n}-q_{n-1}}{\epsilon}-\left(\frac{p_{n}^{2}}{2m}+W_{n}^{h}\right)\right]}\,\,\,\,\,\,\,\,\,\,\,\,\,(\textrm{Discrete }p).\label{eq:full-amp-discrt-p-1}
\end{align}
In both cases above, we have a $\sum_{n}$ in front of the exponential
while we should have an integral instead. In other words, if we wish
to have a normal path integral, we need to pass from a discrete variable,
say $q_{n}$, to the continuous one $q$ (and the same for $p_{n}$).
To achieve this we use the identity \citep{Ashtekar:2010gz} 
\begin{equation}
\sum_{n}\int_{0}^{2\pi}dqf(q,p_{n})e^{ip_{n}q}=\int_{-\infty}^{\infty}dq\int_{-\infty}^{\infty}dpf(p,q)e^{iqp},\label{eq:sum-to-int-q}
\end{equation}
valid for $f(q,p)$ which are periodic in $q$, and a similar expression also holds for the case with discrete $p_{n}$.
Let us consider first the expression (\ref{eq:full-amp-discrt-p-1}).
Using the above identity, one gets for (\ref{eq:full-amp-discrt-p-1})
\begin{equation}
\langle q_{f},t_{f}|q_{i},t_{i}\rangle=\lim_{N\rightarrow\infty,\epsilon\rightarrow0}\left(\prod_{n=1}^{N-1}\int_{-\infty}^{\infty}dq_{n}\right)\left(\frac{\mu}{2\pi\hbar}\prod_{n=1}^{N}\int_{-\infty}^{\infty}dp\right)e^{\frac{i\epsilon}{\hbar}\sum_{n=1}^{N}\left[p_{n}\frac{q_{n}-q_{n-1}}{\epsilon}-\left(\frac{p_{n}^{2}}{2m}+W_{n}^{h}\right)\right]}\,\,\,\,\,\,\,\,\,(\textrm{Discrete }p),\label{C17}
\end{equation}
which is the polymer Feynman formula. Taking the limits in (\ref{C17}) above yields
\begin{equation}
\langle q_{f},t_{f}|q_{i},t_{i}\rangle=\frac{\mu}{2\pi\hbar}\left(\prod_{n}\int_{-\infty}^{\infty}dq_{n}\right)\left(\prod_{n}\int_{-\infty}^{\infty}dp\right)e^{\frac{i}{\hbar}\int dt\left[p\dot{q}-\left(\frac{p^{2}}{2m}+W^{h}\right)\right]}\,\,\,\,\,\,\,\,\,\,\,\,\,\,\,\,\,\,\,\,\,\,\,\,\,\,\,(\textrm{Discrete }p),
\end{equation}
where the effective action can be read off to be
\begin{equation}
S_{\textrm{eff}}=\int dt\left[p\dot{q}-H_{\textrm{eff}}\right]=\int dt\left[p\dot{q}-\left(\frac{p^{2}}{2m}+\frac{\mu^{2}k}{4\hbar^{2}}\csc^{2}\left(\frac{\mu q_{n}}{2\hbar}\right)\right)\right]\,\,\,\,\,\,\,\,\,\,\,\,\,\,\,\,\,\,\,\,\,\,\,\,\,\,\,(\textrm{Discrete }p).
\end{equation}
The case with discrete $q$ is a bit more delicate because of the inherent
discreteness that precludes %of $q$ due to polymer quantization, one can not take
the limit 
\begin{equation}
\lim_{\epsilon\rightarrow0}\left(p_{n}\frac{q_{n}-q_{n-1}}{\epsilon}\right)\rightarrow p\dot{q}.
\end{equation}
 To deal with this, one uses a method similar to Leibniz rule for discrete
variables. One can rewrite the sum that appears in the kinetic term as
\begin{align}
\sum_{n=1}^{N}p_{n}\frac{q_{n}-q_{n-1}}{\epsilon}= & \frac{1}{\epsilon}\left(\sum_{n=1}^{N}p_{n}q_{n}-\sum_{n=1}^{N}p_{n}q_{n-1}\right)\nonumber \\
= & \frac{1}{\epsilon}\left(\sum_{n=1}^{N}p_{n}q_{n}-\sum_{m=0}^{N-1}p_{m+1}q_{m}\right)\nonumber \\
= & \frac{1}{\epsilon}\left(\sum_{n=1}^{N-1}p_{n}q_{n}-\sum_{n=1}^{N-1}p_{n+1}q_{n}+p_{N}q_{N}-p_{1}q_{0}\right)\nonumber \\
= & \frac{1}{\epsilon}\left(-\sum_{n=1}^{N-1}\left[\left(p_{n+1}-p_{n}\right)q_{n}\right]+p_{N}q_{N}-p_{1}q_{0}\right),
\end{align}
which is similar to expressing $p\dot{q}=-q\dot{p}+\dot{(pq)}$ where
$\frac{1}{\epsilon}\left(p_{N}q_{N}-p_{1}q_{0}\right)$ plays the
rule of the ``boundary term'' $(pq)\dot{\,}$ in a discrete sense.
Using this, expression (\ref{eq:full-amp-discrt-q-1}) becomes 
\begin{align}
\langle q_{f},t_{f}|q_{i},t_{i}\rangle= & \lim_{N\rightarrow\infty,\epsilon\rightarrow0}\left(\prod_{n=1}^{N-1}\sum_{n}\right)\left(\left(\frac{\lambda}{2\pi\hbar}\right)^{2}\prod_{n=1}^{N}\int_{-\frac{\pi\hbar}{\lambda}}^{\frac{\pi\hbar}{\lambda}}dp_{n}\right)\times\nonumber \\
 & e^{\frac{i\epsilon}{\hbar}\left[\sum_{n=1}^{N-1}\left[\frac{-\left(p_{n+1}-p_{n}\right)q_{n}+p_{N}q_{N}-p_{1}q_{0}}{\epsilon}\right]-\sum_{n=1}^{N}\left(\frac{2\hbar^{2}}{m\lambda^{2}}\sin^{2}\left(\frac{\lambda p}{2\hbar}\right)+W_{n}^{T}\right)\right]}\,\,\,\,\,\,\,\,\,\,\,\,\,\,\,\,\,\,\,\,\,\,\,\,\,\,\,(\textrm{Discrete }q).
\end{align}
Now, we can take the limit such that the term $\lim_{\epsilon\rightarrow0}\frac{-\left(p_{n+1}-p_{n}\right)q_{n}+p_{N}q_{N}-p_{1}q_{0}}{\epsilon}$
becomes $-q\dot{p}+\dot{(pq)}$, which in the continuous limit can
be rewritten as $p\dot{q}$. Before taking this limit, we use an identity
similar to (\ref{eq:sum-to-int-q}) but for discrete $q$,
\begin{equation}
\sum_{n}\int_{0}^{2\pi}dpf(q_{n},p)e^{iq_{n}p}=\int_{-\infty}^{\infty}dq\int_{-\infty}^{\infty}dpf(p,q)e^{iqp},\label{eq:sum-to-int-p}
\end{equation}
to turn $\sum_{n}$ into an integral. Then by taking the limits, one
gets
\begin{equation}
\langle q_{f},t_{f}|q_{i},t_{i}\rangle=\left(\frac{\lambda}{2\pi\hbar}\right)^{2}\left(\prod_{n}\int_{-\infty}^{\infty}dq_{n}\right)\left(\prod_{n}\int_{-\infty}^{\infty}dp_{n}\right)e^{\frac{i\epsilon}{\hbar}\left[p\dot{q}-\left(\frac{2\hbar^{2}}{m\lambda^{2}}\sin^{2}\left(\frac{\lambda p}{2\hbar}\right)+W^{T}\right)\right]}\,\,\,\,\,\,\,\,\,\,\,\,\,\,\,\,\,\,\,\,\,\,\,\,\,\,\,(\textrm{Discrete }q)
\end{equation}
Note that the potential does not have a discrete subscript anymore
since we turned $q_{n}$ into a continuous variable $q$ by using
(\ref{eq:sum-to-int-p}). Thus the effective action can be read off
to be 
\begin{equation}
S_{\textrm{eff}}=\int dt\left[p\dot{q}-H_{\textrm{eff}}\right]=\int dt\left[p\dot{q}-\left(\frac{2\hbar^{2}}{m\lambda^{2}}\sin^{2}\left(\frac{\lambda p}{2\hbar}\right)+\frac{k\hbar^{4}}{\lambda^{4}}\left(\sqrt{|q-\lambda|}-\sqrt{|q+\lambda|}\right)^{4}\right)\right]\,\,\,\,\,\,\,\,\,\,\,\,\,\,\,\,\,\,\,\,\,\,(\textrm{Discrete }q).
\end{equation}

\section{Action of $\hat{H}_{E}$ in $(\hat{q},\hat{V}_{\lambda})$ polarization\label{sec-app:rep-HE-q,V}}

\subsection{In $|q_{n}\rangle$ basis\label{sub-app:rep-HE-q,V-base-q}}

The action of the Hamiltonian (\ref{eq:Euclid-H-Qreps}) on the basis
$|q_{n}\rangle$ in the polarization (\ref{eq:q-descrt-polarz-1})-(\ref{eq:q-descrt-polarz-3})
can be computed as follows. The kinetic term acts like 
\begin{equation}
\frac{\widehat{p_{\lambda}^{2}}}{2m}|q_{n}\rangle=\frac{\hbar^{2}}{2m}\left(\frac{2-\hat{V}_{\lambda}-\hat{V}_{-\lambda}}{\lambda^{2}}\right)|q_{n}\rangle=\frac{\hbar^{2}}{2m\lambda^{2}}\left(2|q_{n}\rangle-|q_{n}-\lambda\rangle-|q_{n}+\lambda\rangle\right),
\end{equation}
where the operator $\widehat{p_{\lambda}^{2}}$ is defined in (\ref{eq:p2lambda-op-def}).
The potential $\hat{W}$ in (\ref{eq:Euclid-H-Qreps}) has two terms
for each of which we have
\begin{equation}
\frac{\hat{V}_{-\lambda}}{\lambda}\left[\widehat{\sqrt{|q|}},\hat{V}_{\lambda}\right]|q_{n}\rangle=\frac{1}{\lambda}\sqrt{|q_{n}-\lambda|}|q_{n}\rangle-\frac{1}{\lambda}\sqrt{|q_{n}|}|q_{n}\rangle
\end{equation}
and
\begin{equation}
\left[\widehat{\sqrt{|q|}},\hat{V}_{\lambda}\right]\frac{\hat{V}_{-\lambda}}{\lambda}|q_{n}\rangle=\frac{1}{\lambda}\sqrt{|q_{n}|}|q_{n}\rangle-\frac{1}{\lambda}\sqrt{|q_{n}+\lambda|}|q_{n}\rangle.
\end{equation}
Thus the full expression for the action of the potential in this basis
becomes
\begin{equation}
\hat{W}|q\rangle=k\hbar^{4}\left(\frac{\hat{V}_{-\lambda}}{\lambda}\left[\widehat{\sqrt{|q|}},\hat{V}_{\lambda}\right]+\left[\widehat{\sqrt{|q|}},\hat{V}_{\lambda}\right]\frac{\hat{V}_{-\lambda}}{\lambda}\right)^{4}|q_{n}\rangle=\frac{k\hbar^{4}}{\lambda^{4}}\left(\sqrt{|q_{n}-\lambda|}-\sqrt{|q_{n}+\lambda|}\right)^{4}|q_{n}\rangle
\end{equation}

\subsection{In $|p\rangle$ basis\label{sub-app:rep-HE-q,V-base-p}}

One can act the Hamiltonian (\ref{eq:Euclid-H-Qreps}) on basis $|p\rangle$
where
\begin{equation}
\hat{V}_{\lambda}|p\rangle=e^{\frac{i}{\hbar}\lambda p}|p\rangle.
\end{equation}
Then the kinetic term in (\ref{eq:Euclid-H-Qreps}) turns out to be
\begin{align}
\frac{\widehat{p_{\lambda}^{2}}}{2m}|p\rangle=\frac{\hbar^{2}}{2m}\left(\frac{2-\hat{V}_{\lambda}-\hat{V}_{-\lambda}}{\lambda^{2}}\right)|p\rangle= & \frac{2\hbar^{2}}{m\lambda^{2}}\sin^{2}\left(\frac{\lambda p}{2\hbar}\right)|p\rangle\label{eq:p2-poly-sin2}
\end{align}
where we have used (\ref{eq:p2lambda-op-def}) and (\ref{eq:p2lambda-op-act}).
To find the action of the potential term, we first note that
\begin{align}
\hbar^{4}\left(\frac{\hat{V}_{-\lambda}}{\lambda}\left[\widehat{\sqrt{|q|}},\hat{V}_{\lambda}\right]+\left[\widehat{\sqrt{|q|}},\hat{V}_{\lambda}\right]\frac{\hat{V}_{-\lambda}}{\lambda}\right)^{4}|p\rangle= & k\hbar^{4}\left(\frac{\hat{V}_{-\lambda}}{\lambda}\left[\widehat{\sqrt{|q|}},\hat{V}_{\lambda}\right]+\left[\widehat{\sqrt{|q|}},\hat{V}_{\lambda}\right]\frac{\hat{V}_{-\lambda}}{\lambda}\right)^{4}\sum_{n}|q_{n}\rangle\langle q_{n}|p\rangle.
\end{align}
Then using the results of the previous subsection, we get for the
action of the potential in this case
\begin{equation}
k\hbar^{4}\left(\frac{\hat{V}_{-\lambda}}{\lambda}\left[\widehat{\sqrt{|q|}},\hat{V}_{\lambda}\right]+\left[\widehat{\sqrt{|q|}},\hat{V}_{\lambda}\right]\frac{\hat{V}_{-\lambda}}{\lambda}\right)^{4}|p\rangle=\frac{k\hbar^{4}}{\lambda^{4}}\sum_{n}\left(\sqrt{|q_{n}-\lambda|}-\sqrt{|q_{n}+\lambda|}\right)^{4}|q_{n}\rangle\langle q_{n}|p\rangle.
\end{equation}

\bibliography{Poly-bib}

\end{document}